# Hardware Accelerated SDR Platform for Adaptive Air Interfaces

*Tarik Kazaz, Christophe Van Praet, Merima Kulin, Pieter Willemen, Ingrid Moerman*
*(Department of Information Technology (INTEC), Ghent University - iMinds)*

*Abstract*—The future 5G wireless infrastructure will support any-to-any connectivity between densely deployed smart objects that form the emerging paradigm known as the Internet of Everything (IoE). Compared to traditional wireless networks that enable communication between devices using a single technology, 5G networks will need to support seamless connectivity between heterogeneous wireless objects and IoE networks. To tackle the complexity and versatility of future IoE networks, 5G will need to guarantee optimal usage of both spectrum and energy resources and further support technology-agnostic connectivity between objects. One way to realize this is to combine intelligent network control with adaptive software-defined air interfaces. In this paper, a flexible and compact platform is proposed for on-the-fly composition of low-power adaptive air interfaces, based on hardware/software co-processing. Compared to traditional Software Defined Radio (SDR) systems that perform computationally-intensive signal processing algorithms in software, consume significantly power and have a large form factor, the proposed platform uses modern hybrid FPGA technology combined with novel ideas such as RF Network-on-Chip (RFNoC) and partial reconfiguration. The resulting system enables composition of reconfigurable air interfaces based on hardware/software co-processing on a single chip, allowing high processing throughput, at a smaller form factor and reduced power consumption.

*Keywords—Software-Defined Radio; Software-Defined Wireless Networks; Wireless Networks; Reconfigurable computing*

## I. INTRODUCTION

Currently deployed wireless infrastructures are based on inflexible and closed processing platforms. Due to lack of flexibility and programmability, such infrastructures would imply a 10-year cycle for the standardization and massive deployment of next-generation wireless network technologies. At the same time, realizing innovations from the wireless domain in real-world scenarios is practically unfeasible. On the other side, a more programmable infrastructure could facilitate the application of innovations developed by the research community, in real life scenarios such as context-aware cross-layer optimization of network parameters [7]. In this way, recent research achievements on improving wireless networks could be evaluated and implemented on the fly, and the long cycle for applying changes in wireless infrastructures would be avoided.

One of the key architectural principles of the vision of 5G wireless networks is to support efficient and flexible usage of the wireless spectrum and radio access technologies, and easy adoption of innovations as they appear. Industry and standardization bodies are in the early phase of the definition of 5G technologies, and major demands that are shaping design are already known, such as heterogeneous connectivity and 1000-fold data traffic increase. To meet such challenging demands, two approaches for enabling the implementation of air interfaces for 5G Networks are proposed. The first approach utilizes massive MIMO techniques to achieve higher spatial gains and increase the spectral efficiency. The second approach exploits hyper-dense deployments of heterogeneous and small cell networks (HetSNets) [1] located close to wireless users to cope with the increasing traffic demands in indoor environments. HetSNets could support the offloading of outdoor traffic in cases when mobile users are close to small cells. These two approaches need to be supported by an appropriate architecture. The most promising solution so far is the cloudification of wireless networks as proposed by the Cloud RAN (C-RAN) architecture [3]. The C-RAN architecture is based on centralized processing of baseband signals (I-Q samples) in data centers, while distributed radio heads are used for RF processing and the transmission of signals. This architecture decreases the operational cost by replacing the processing done by multiple base stations with one processing unit, and enables implementation of new techniques for joint processing and demodulation of signals from multiple users. It also simplifies the implementation of techniques for interference mitigation in both, HetSNets [1] and macro cells, such as coordinated multipoint (CoMP) [5] and enhanced inter-cell interference coordination (eICIC) [6]. However, the C-RAN architecture requires that digital I-Q samples are transported between multiple remote radio heads (RRHs) and the centralized processing unit, which demands extremely high data rate and low latency connection between them. To address this problem, more distributed baseband signal processing can be applied instead of a fully centralized processing approach [8].

In this paper we argue that the infrastructure of future 5G wireless networks requires flexible and programmable components as its main building blocks, in order to cope with rapidly changing wireless contexts and accelerate the deployment of innovations. To reach this goal, concepts from Software Defined Radio (SDR) and Software Defined Networking (SDN) can be exploited. SDN is based on the standardized OpenFlow [9] protocol that enables programmability of wired networked devices at the network layer. This concept is already accepted by the industry and is deployed in many campus and data center networks, which enables a faster innovation cycle in wired networks [8, 29]. More recently, the SDR approach introduces similar principles and goals for wireless networks. However, this is still a challenging task due to the very different nature of wireless network technologies each having its own specifics, often leading to implementations of non-programmable logical circuits on Application Specific Integrated Circuit (ASIC). ASIC implementations have shown to be optimal solutions for wireless standards due to its high efficiency in terms of power consumption and computation performance, especially for battery powered mobile devices. At the same time, ASICs are static and therefore not suitable for SDR. The flexibility requirement makes general purpose processors (GPP) as a preferable hardware platform for implementing SDR systems. Due to their sequential nature of processing, GPPs are not suited for high throughput computing with real-time requirements. In order to overcome issues related to the processing power of GPP based solutions, platforms based on graphical processing units (GPUs) have emerged. These platforms offer more processing power while preserving the same flexibility as GPPs, at the price of higher power consumption and hence, lower energy efficiency.

State-of-the-art platforms for implementing SDR systems require the setup of a fully-functional desktop computer that consumes significant power and has a large form factor, which precludes any real-life deployment. Traditional computing platforms like GPPs and GPUs do not scale anymore in accordance with Koomey's Law [10] for computing efficiency. For this reason, traditional GPU vendors recognized the need for a radical shift of the traditional computing approach, and they are developing new computing platforms which couple processors together with FPGAs [11][12]. Hybrid platforms that enable computing based on a combined software and hardware approach are expected to overcome the stalled performance scaling and improve energy efficiency.

This paper is organized as follows: First, we present an overall logical architecture and initial implementation of a programmable radio data plane with real-time performance. Following the emergence of new computational platforms, the proposed system is based on a hybrid FPGA chip. The system aims to ensure fulfilment of three key objectives: real-time performance, flexibility and scalability (Section II). Real-time performance is essential for implementing wireless protocols that can meet the requirements as defined in standards. This condition was not fulfilled with the traditional SDR approach such as used in GNU Radio [13], based on pure software processing on host commodity computer units. We propose a hardware accelerated SDR approach in order to ensure ASIC-like processing capabilities while preserving flexibility. In order to ensure flexible control and adaptability to new evolutions, our architecture follows the principles of data and control plane separation (Section III). Although SDR and SDN are discussed as separated paradigms, the proposed architecture benefits from both concepts by extending the network data plane, as considered today in SDN, with a radio data plane for wireless networks.

In this paper we present a first step towards the implementation of the proposed real-time flexible and scalable architecture on a hybrid FPGA platform (Section IV). More specifically, this papers presents the first results of the implementation of a programmable and controllable radio data plane with real-time performance on a Zynq-based FPGA platform. Finally, we discuss the main challenges that need to be addressed for the system realization and motivate its usability with potential use cases. We conclude the paper in Section VIII.

## II. DESIGN GOALS AND PRINCIPLES

Ideal SDR platforms support the implementation of flexible and upgradeable air interfaces with real-time performance requirements. To meet such objectives, SDR platforms need to ensure two key requirements related to processing and reconfiguration capabilities. *Real-time performance* is related to the completion of processing tasks within an exactly set time frame. For instance, IEEE 802.11n technology, which has a relatively wide channel bandwidth compared to other technologies, is an interesting example due to processing constraints that it implies. The worst-case scenario assumes to analyze the 40MHz bandwidth mode, which assumes double channel (I-Q) sampling of at least 80Msps. From a processing perspective, this implies that the platform is able to support processing of two streams of samples with data rate of 80Msps each. If the SDR platform fulfills such a hard constraint on processing, then the implementation of other wireless technologies such as 3GPP LTE-Advanced, IEEE 802.11a/g/p/ah and IEEE 802.15.4 with narrower channel bandwidths can also be easily supported.

The air interface of a typical wireless standard can be abstracted as a chain of processing units that are performing signal processing functions such as filtering, coding or Fast Fourier Transformation (FFT). In order to support, cross-layer real-time performance the platform should enable low latency interfaces between the medium access control (MAC) layer and the processing chains. In this way, time critical operations that are defined in the standard of a specific MAC layer protocol, can be executed on time (e.g. fast acknowledgment of a packet reception).

To ensure *reconfiguration* capabilities, the SDR platform should provide interfaces for control and full reconfiguration of air interfaces. Control interfaces are providing mechanisms for real-time adaption of processing chains, when there is need to make smaller modifications of currently deployed processing chains. A good example of such a modification, is the change of modulation and coding schemes in the IEEE 802.11n technology. Full reconfiguration of processing chains enables the ability to deploy completely new processing chains on the SDR platform. Such functionality can be used in cases when there is a need to do extensive changes in processing chain which cannot be accomplished by using only the control interface. For example, switching between two wireless technologies which have completely different processing chains such as IEEE 802.15.4 and IEEE 802.11g can be an example of such a scenario. Moreover, parallel deployment of more than one processing chain which corresponds to different wireless technologies should be supported as long as there are enough hardware resources on the platform itself.

## III. OVERVIEW

The architecture of a hardware accelerated SDR platform is inspired by the general idea of Software Defined Networking (SDN) originating from the domain of wired networks. However, wireless technologies and protocols are much more diverse making it challenging to abstract the physical layer of each wireless technology in a unified way. In order to achieve the design goals defined in Section II, during the design phase, a distinct separation between the data and control plane has to be defined. Fig. 1. shows the high-level architecture of our system, and illustrates the logical separation between the data and control plane interfaces. The data plane is an open and programmable data processing infrastructure, which consists of a network data plane (the software defined network data plane that operates as a network forwarding infrastructure) and a radio data plane (a software defined radio data plane that operates as a signal processing infrastructure, RF infrastructure and a medium access control infrastructure). This definition is an extension of the typical data plane definition from the SDN community, which considers only wired networks as a programmable infrastructure and network forwarding as the only operation over data (packets). In addition, we extend the term data by providing abstraction over three forms of data in the communication networks: data packets, signal samples and radio waves.

The network data plane is comprised of forwarding engines, whose capabilities are exposed towards the control plane via the network control interface. The network control interface is equivalent to the Control-Data Plane Interface (CDPI) in SDN networks [14]. CDPI provides: (i) programmatic control of all forwarding operations, (ii) capabilities advertisement, (iii) statistics reporting, and (iv) event notification. This provides seamless incorporation of OpenFlow protocol into our architecture.

The radio data plane consists of reconfigurable and programmable radio processing chains, which exposes control and processing capabilities through the radio control interface. Based on its functionalities it can be separated into three logical sub-planes: the medium access plane, the radio processing plane and the RF plane.

The medium access plane enables implementation of various channel access schemas and performs MAC layer related framing. The basic functional unit of the medium access plane is the MAC processing engine [15, 16], which is performing programmable scheduling of packet transmissions.

The radio processing plane is performing baseband processing functionalities through the usage of processing chains. To fulfill the flexibility requirements, the proposed system architecture follows the principle of decomposition of the processing chain [17]. A single processing chain can be decomposed into a group of processing units that are representing specific signal processing blocks. The decomposition of processing chains into processing units enables reusability of common processing units between several processing chains. For example, one processing chain can represent the whole receiver PHY layer of a specific wireless standard such as, for instance, IEEE 802.11g or 3GPP LTE.

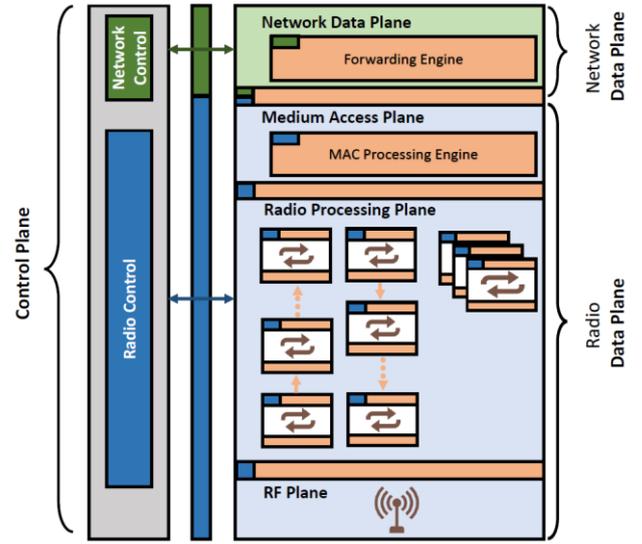

Fig. 1. High-level architecture of a hardware accelerated SDR platform

Those two particular technologies are both based on OFDM communications technology, meaning that there is a possibility to share a number of common processing units in their processing chains such as: FFT, Channel Estimation, QAM mapping/de-mapping, etc. One aspect of sharing resources is component reusability, which aims to enable flexible modification of processing chains, e.g. the composition of LTE PHY layer on the basis of the IEEE 802.11g PHY layer. Another case of sharing resources aims to increase utilization of FPGAs in situations when there is a need to have parallel chains, while at same time already deployed processing units are able to process data several times faster than the data rate itself. Hybrid FPGAs like Xilinx Zynq contain FPGA fabric, which can meet timing constraints at clock rates up to 300 MHz [18], while data rates of oversampled I-Q signal stream for major wireless technologies do not exceed 80Msps (except IEEE 802.11ac for channels of 80MHz and 160MHz). In those situations, it is possible to use the technique of time-multiplexed sharing of processing units between chains. However, sharing of resources between processing chains could be used only in situations when latency and computational throughput requirements of processing chains stay in accordance with the applied wireless standard(s). As an example, consider the IEEE 802.11g standard that specifies the short inter-frame space (SIFS) of 10μs as the time interval within which IEEE 802.11g compliant device need to acknowledge the reception of a packet after receiving the last symbol of it. Such a strong constraint on latency can be violated if time-multiplexed sharing of resources is not applied properly. The radio control interface for the radio processing plane enables control and reconfiguration capabilities towards the control plane. The *radio control interface* is a logical abstraction over two physical interfaces: the interface for parametrized control of processing units and the interface for reconfiguration of processing chains. The *parametrized control interface* enables minor real-time modifications of the behavior of processing units, such as changing the FFT-length of the FFT block or changing the rate of the coder. The radio reconfiguration interface enables full reconfiguration of processing chains on FPGA fabric. In case when there is need for extensive modifications of processing chains such as swapping from one wireless technology to another, the radio control plane accomplishes this through the *radio reconfiguration interface*. Ideally, reconfiguration will target processing units or parts of processing chains, which should be swapped. However, this fine granularity of reconfiguration on FPGAs can be accomplished just through the usage of partial reconfiguration techniques of FPGA fabric. Potential challenges of partial reconfiguration will be examined in Section V.

The *radio frequency (RF) plane* performs analog radio frequency signal processing. This plane provides a control interface towards the control plane, to enable control over RF processing blocks, such as control of: the frequency of the local oscillator (carrier frequency-channel selection), the bandwidth of the analog filter, the gain of the power amplifier or gain of the low noise amplifier, etc.

The control plane consists of radio controllers and network controllers, and it provides an abstract view of network to network management tools and customized control applications. Network controller follows OpenFlow paradigm and it enables incorporation of OpenFlow into wireless devices. Because wireless networks are fundamentally different compared to wired networks, current capabilities of OpenFlow are not sufficient for control of medium access and radio parameters. In order to enable control of the radio data plane, control plane needs to be extended with radio controllers which provide control and an abstract view over radio data plane. This is addressed by ongoing project WiSHFUL [19].

## IV. INITIAL IMPLEMENTATION

The high-level architecture of the hardware accelerated SDR platform, as presented in Fig. 1, is oriented towards the exploitation of reconfiguration and processing capabilities of hybrid FPGAs for the implementation of adaptive air interfaces. Hence, this paper focuses on aspects that are related to the radio data plane implementation. However, our final goal is to design and develop a platform that enables the implementation of programmable radio and network data planes. In order to understand how the logical architecture of our system maps to physical resources, we analyze the generic hardware architecture of a hybrid FPGA chip that is used for the implementation.

### A. Hardware Architecture of Hybrid FPGAs

Hybrid FPGAs present a compact and flexible architecture that tightly couples a hardware processing system (PS) and a programmable logic (PL) by utilizing high throughput interfaces and general purpose interfaces in between. The invention of these hardware platforms provides the necessary precondition for the implementation of efficient hardware-software systems, which can benefit from both (i) parallelized and deterministic computational capabilities of the FPGA fabric and (ii) flexibility of hardware processors. For the study, a Xilinx Zynq 7045 chip was used, which contains a dual-core ARM Cortex-A9 processor as the PS unit and a Kintex-7 FPGA as the PL unit. Fig. 2 shows a simplified block diagram of a Xilinx Zynq architecture.

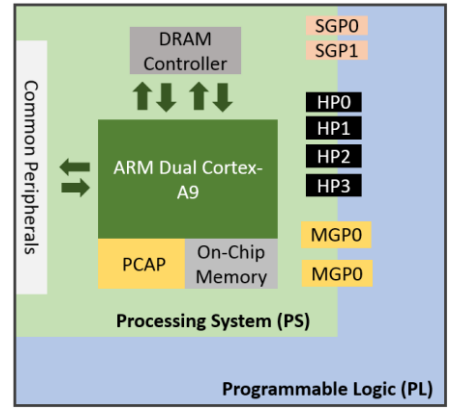

Fig. 2. Zynq Architecture showing PS, PL and the interfaces

The PL unit, Kintex-7 FPGA, offers 350 K programmable logic cells and 900 programmable embedded DSP slices. The basic PL circuit resources on this chip can meet timing constraints at clock rates up to 300 MHz [18] which provides high processing capabilities needed to perform computationally intensive tasks. The PS unit, dual core ARM processor, offers a maximum clocking rate of 866MHz and is able to run stand-alone operating systems such as e.g. Linux Ubuntu.

Between the PS and the PL logic, four high-performance (HP) interfaces, HP0-HP3, offer high-rate data transfers with a maximum total throughput of 9.6GByte/s [18]. This throughput can be flexibly allocated between PS and PL depending on the needs and the design of the target implementation. Besides the HP interfaces, there are a number of general purpose (GP) memory mapped interfaces which allow bidirectional register access between PS and PL.

### B. Medium Access Plane

The medium access plane is fully implemented in software on the PS unit of the Zynq chip. It interacts with the radio processing plane through the HP interfaces. Two software ring buffers are implemented to load packets from/to receiver/transmitter processing chains. The data transfer between the radio processing plane and the medium access plane that goes through HP interfaces is handled by the Direct Memory Access (DMA) controllers.

The HP interfaces are implemented in accordance with the AMBA AXI standard [20] for on chip communication. There are three types of AXI interfaces described in the current AXI4 standard [20]: (i) AXI4 – high throughput memory mapped interface, (ii) AXI Lite – simplified AXI4 interface for low throughput memory mapped interface, (iii) AXI4 Stream – high throughput streaming data interface. In order to reduce the signaling overhead between the medium access and processing plane, AXI4 Stream is chosen for flow-controlled data transfers. Since the data is in form of packets with arbitrary size, the most appropriate choice for data transfer is the AXI4 Stream interface in burst mode.

The DMA controls the data transfer using two function calls:
- *send_packet(addr_read, packet_size)*,
- *receive_packet(addr_write, packet_size);*

and two interrupts:
- *sent_IRQ*,
- *received_IRQ*.

The corresponding function parameters are:
- *addr_read* – the memory address where a packet is stored,
- *addr_write* – the memory address where a received packet will be stored,
- *packet_size* – the length of a packet expressed in bytes.

The interrupt *sent_IRQ* will be generated in case the transmission of a packet from PS to PL is completed, while for the opposite direction the *received_IRQ* will be generated.

*C. Radio Processing Plane*

Fig. 3. presents how the logical architecture shown on Fig. 1. maps to physical resources on the hybrid FPGA chip. The radio processing plane consists of many processing units that are implemented as hardware accelerators on the PL side of the Zynq chip. Every processing unit is characterized by the number of input samples, the number of output samples and its computational throughput.

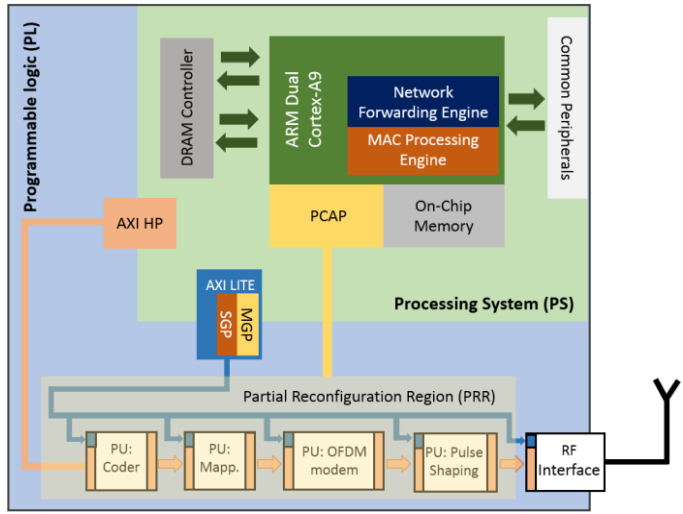

Fig. 3. System architecture of SDR platform based on a hybrid FPGA

Processing units are the building blocks of processing chains. To enable control of the processing units, the interconnection between them, and their flexible arrangement into processing chains, a standardized physical interfaces is needed. In addition, in order to prevent bottle-necks in the processing chains, it is important that the processing units are implemented so that they provide sufficient computational throughput. The radio processing plane provides real-time flexibility through reconfiguration at two levels: the processing units and the processing chains.

The run-time adaptation of processing units is provided through their parametrization (parametric reconfiguration). Parametrized processing units contain several control registers, which are exposed to the control plane through AXI4 memory-mapped interfaces. Parametrized processing units can be modified without the need to reprogram the PL logic and generate new bitstreams, as they are implemented physically on the PL logic. A typical example for applying parametric reconfiguration may be the control of the FFT-length on a FFT processing unit, or the control of the CRC-polynomial on a CRC processing unit. However, parametric reconfiguration may not be sufficient to support modifications of the PL logic at the level of processing chains, such as completely changing one or more processing units in a processing chain.

The reconfiguration of processing chains may be realized through switching between different FPGA bitstreams that are generated for each processing chain and pre-stored in memory. The reconfiguration process is performed by the application that is running on PS unit through the Processor Configuration Access Port (PCAP) and its controller that provide a reconfiguration throughput of 128MB/s [21]. The disadvantage of this reconfiguration approach is that it changes the entire PL logic, even parts that do not require modifications. For instance, to add or replace only a single processing unit of a processing chain, all existing processing units need to be reconfigured, even those that do not require modifications. A more efficient approach for the reconfiguration of processing chains is partial reconfiguration. Partial reconfiguration enables finer reconfiguration granularity, by enabling reconfiguration of a part of the PL logic instead of the whole PL logic. This approach requires the definition of partially reconfigurable regions (PRRs), which are able to host parts or whole processing chains. Reconfiguration of PRRs requires generation of partial bitstream files, which contain all the necessary configuration information to place new processing units in the right PRR of the processing chain.

Systems for wireless signal processing are usually implemented using a streaming model for computing on FPGAs. The streaming model of computing is characterized by streams of data passing through processing units [22]. Based on our previous discussion about AXI interfaces it is obvious that AXI4-Stream is the natural choice to implement interfaces between processing units so that streaming of data samples through them is possible as well as the composition of processing chains. Basic signaling between processing units is based on a T_READY/T_VALID handshake, which assumes that (i) the receive processing unit asserts a T_READY signal in the AXI Stream bus when it becomes ready to receive and process data, and (ii) the sender processing unit asserts a T_VALID signal when it sends valid chunks of data. However, this is not enough to enable asynchronous design and sharing of processing units between processing chains. Asynchronous design needs to guarantee that every processing unit or part of a processing chain can exist in its own clock domain while preserving the ability to interact with other processing units. Communication between interacting processing units is provided through a Network-on-Chip (NoC) architecture. NoC is a promising architecture for emerging System-on-Chip (SoC) applications on FPGAs [23]. In particular, our system uses a RFNoC [24] architecture to interconnect shared processing units with the other part of the according processing chains. The RFNoC architecture is based on a crossbar switch, which is able to route streams of data that are packed in accordance with the compressed header (CHDR) format. We claim that there is no need to interconnect processing units that are not shared between processing chains through the RFNoC switch, as it increases consumption of PL logic resources and introduces overhead. In order to support both direct interconnections and through RFNoC switch interconnections between processing units we use RFNoC interface wrappers just for the processing units that will be connected to the RFNoC switch. This enables flexible refactoring of processing units interfaces in order to adopt the way they interact with each other and with the rest of the system. Incorporation of RFNoC architecture with standardized CHDR header format supports scalable extension of processing chains in cases when there is more hardware processing power needed. For external communication between hardware units, the RFNoC switch encapsulates CHDR frames into VITA-49 Radio Transport (VRT) [25] standardized format. This provides the possibility to interconnect several Zynq SoCs into processing clusters through 10 Gigabit Ethernet connections, and enables transparent connection between processing units which are located on different physical chips. For example if hardware resources on a single chip are not sufficient to implement a certain processing chain, several chips could be interconnected to enable distributed implementation of the processing chain. This can be useful for massive MIMO systems as they require independent processing of a number of I-Q data streams, which is very computation-intensive.

*D. RF Plane*

The radio frequency (RF) plane may consist of Analog Devices AD-FMCOMMS1 or AD-FMCOMMS2 RF boards that are connected through FMC connectors. FMC connectors provide a high throughput interconnection between the FPGA and the ADC and DAC converters on the RF boards. The RF plane is controlled over the radio control plane, which enables the control of RF parameters such as the frequency of the local oscillator or the gain of the low noise amplifier (LNA).

V. CHALLENGES

The key challenges for the implementation of SDR platforms on Hybrid FPGAs are (i) the implementation of a standardized interfaces for the control of processing units and the communication between them, and (ii) fine-grained reconfiguration of the FPGA logic. To address the first challenge standardized protocols for on chip communication like AMBA AXI protocol [20] and RFNoC concepts, may be utilized. FPGA reconfiguration requires that bitstream files are generated and accessible to the reconfiguration controller. Fast generation of bitstream files for the reconfiguration of the FPGA logic, as well as efficient reconfiguration of FPGAs are remaining as open issue. Proprietary tools for FPGA design offer slow compilation flow for generation of FPGA bitstream files. This is a major obstacle for fast composition and assembly of FPGA designs. Utilization of alternative FPGA compilation flows, as used in TFlow [37] would overcome this challenge, by enabling on the fly generation of bitstreams based on the library of precompiled modules. With TFlow our platform could enable on the fly composition of processing chains formed from processing units that would be stored in repositories. Fine granular re-programmability of the FPGA logic can be done only by applying a partial reconfiguration technique. However, efficient partial reconfiguration requires that physical FPGA resources are split into partially reconfigurable fixed-sized regions (PRRs), which precludes the dynamic allocation of FPGA resources. Specifically, for our proposed platform, partial reconfiguration would target processing units which requires segmentation of the FPGA logic in many fixed-sized regions. Such fine-grained reconfiguration of the FPGA logic is difficult to achieve with current design tools. For this reason, we are considering to apply this technique for the reconfiguration of processing chains.

VI. USE CASES

In this section we introduce a few use-cases to illustrate some benefits that the programmable wireless infrastructure can introduce for wireless networks.

1. *Fine grade base station decomposition for C-RAN:* The initially idea of C-RAN focuses on coarse-grained function split between RRHs and centralized baseband processing units (BBU), where the entire baseband signal processing is located at the BBU. This requires digital I-Q samples to be transported between multiple remote radio heads (RRHs) and the centralized processing unit, which demands extremely high data rates and low latency connections between them. As an example, we consider the transmission of LTE signals over the standardized interface for fronthaul networks called CPRI (Common Public Radio Interface) [26]. In case of the RRH has eight antennas and support for 20 MHz LTE channels, a transmission rate of 7.36 Gbps will be required, which is close to maximum capacity of single CPRI link. In general, the requirements on the transmission rate between RRHs and the centralized processing unit are increasing linearly with the number of antennas and the channel bandwidth, which is not in line with implementation of massive MIMO technologies. To address this problem the authors of [8] introduce fine-grained base station decomposition by leaving partial baseband processing at RRHs. However, the authors do not discuss how such architecture could be implemented. A programmable radio data plane based on the proposed hardware accelerated SDR platform could be a viable solution, and enable real-time performance and programmability of remote and central baseband processing units being both part of same radio processing data plane.

2. *Programmable multi radio IoT gateways:* Emerging paradigm known as the Internet of Things (IoT) requires support for connectivity between heterogeneous wireless objects. In order to support such diversity of wireless technologies there is need for deployment of devices which have multi radio capabilities. The proposed hardware accelerated SDR platform can facilitate the implementation of such devices with additional support for deployment of new wireless technologies in future. A new technology may be deployed as a combined package for reprogramming of both radio and network data planes

3. *Context-Aware wireless link adaption:* Ultra-dense deployment of heterogeneous wireless devices will require efficient utilization of spectrum and energy resources. One approach that can enable efficient utilization of spectrum and energy resources is context-aware adaptation of wireless link parameters [7]. Context-aware wireless systems are able to dynamically adjust physical layer parameters in order to efficiently utilize available energy and spectrum resources while maintaining the desired quality of communication. For example, incorporation of this functionality in Wi-Fi networks would enable adjustment of channel width and modulation and coding schemes based on service requirements. Interfaces for parametrized control of radio data plane that are part of our architecture can support those functionalities.

## VII. RELATED WORK

The key idea of flexible wireless systems relies on concepts introduced by SDR [27], which assumes shifting from hardware design approach towards flexible system where major components are implemented in software. The absence of appropriate platforms that could provide sufficient processing power within small energy consumption footprint precluded massive deployment of reliable SDR-based wireless systems in real-life scenarios. As a result, the idea of SDR is related to the research prototyping and the hobbyist studies. At the same time it is predicted that future wireless system will need to support a tremendous 1000-fold increase of data traffic in a decade. The spectral efficiency of point-to-point technologies are very close to the Shannon theoretical limit, and it became obvious that future wireless systems will need to support a more flexible and coordinated approach for efficient utilization of limited spectrum resources [1, 2]. Two key enablers, HetSNets and C-RAN, already introduced in Section I, both require a flexible and programmable infrastructure based on the concepts of SDR and SDN [28]. The integration of SDR and SDN approaches is forming new paradigm, which is referred to as Software-Defined Wireless Networking.

Recently, few architectures are proposed [29, 30, 31, 8, 3] for future wireless network organization. SoftRAN and V-Cell [30, 31] are introducing a centralized control plane for radio access networks by abstracting radio infrastructure as big-base station with an aggregated pool of radio resources. Cloud RAN [3] architecture proposes centralized baseband processing for wireless networks in order to support flexible control and management of processing and energy resources. In addition BigStation and CloudIQ [32, 33] are evaluating the benefits of the C-RAN approach through processing of LTE signals on a centralized GPP-based infrastructure. Besides other observations, both approaches came to the same conclusion: the GPP-based infrastructure cannot meet current requirements in terms of performance-per-Watt. OpenSWDN [29] introduces a novel Wi-Fi architecture based on an SDN/NFV approach, and it enables programmable control and virtualization of Wi-Fi physical resources through the usage of light virtual access point (LVAP). However it is based on nonprogrammable wireless cards and it does not exploit the benefits of an SDR approach. Several solutions are proposed for implementation of programmable radio data paths based on the SDR approach [34, 35, 36]. OpenRadio and Atomix [34, 35] share the same idea, which is a framework for the implementation of programmable radio data planes on multi-core DSP chips. They introduce a novel way for mapping signal processing applications on a DSP architecture, which is formed from several processing cores and hardware accelerators. Due to the sequential nature of the processing cores that are available on the DSP chip, the implementation of an IEEE 802.11g receiver in the Atomix framework can achieve 36ms processing latency with 1.5ms of variability, which is still sufficient to meet the requirements of a 5MHz IEEE 802.11a channel, but not anymore for the case of 10MHz and 20MHz. A novel platform for the implementation of flexible radio data paths on hybrid FPGAs is presented in [36]. However, their work is mainly concentrated on the utilization of partial reconfiguration techniques for real-time adaptation of radio data paths. TFlow [37] introduces an alternative FPGA compilation flow to reduce back-end time required to generate bitstream files. With TFlow our platform could enable on the fly composition of processing chains based from processing units that would be stored in repositories. Our work merges the ideas from the SDR [34, 35, 36] and SDN [8, 29] domain in order to provide a platform for the implementation of fully controllable and programmable radio and network data paths.

## VIII. CONCLUSION

Due to lack of flexibility and programmability, current wireless infrastructures imply a 10-year cycle for the standardization and massive deployment of next-generation wireless network technologies. At the same time, current SDR hardware platforms do not provide the necessary computational power for massive deployments of SDR-based wireless infrastructures. The invention of hybrid computing platforms, which combine the traditional GPP-like computing approach with the FPGA-based hardware acceleration approach, shows promise for the implementation of standalone energy efficient SDR-based adaptive air interfaces.

In this paper, we proposed a novel SDR-based system for on-the-fly radio reconfiguration. The proposed implementation makes use of recent advances in the field of FPGA-accelerated design and GPP computing, providing a highly scalable and flexible solution for real-time signal processing. Further, we proposed a high-level architecture for re-programmability of the radio and network plane, discuss the implementation of SDR platforms on top of Hybrid FPGAs and demonstrate the position of the proposed system within the overall architecture. Finally we detail the implementation of our hardware accelerated SDR platform.

ACKNOWLEDGMENT

The research leading to these results has received funding from iMinds Strategic Research Program on Internet of Things, the European Horizon 2020 Programmes under grant agreement n°645274 (WiSHFUL project) and n°671563 (Flex5Gware project), and the SBO SAMURAI project (Software Architecture and Modules for Unified RAdIo control).

ABOUT THE AUTHORS

**Tarik Kazaz** received his M.Sc. degree with honors in Electrical Engineering from the University of Sarajevo, in 2012. In 2013 he joined BH Mobile, where he was working as Radio Access Network Engineer, while at the same time he was part time teaching assistant at Faculty of Electrical Engineering, University of Sarajevo. Since January 2015, he is PhD student at the department of Information Technology (INTEC) at Ghent University. Within this department, he is working at the Internet Based Communication Networks and Services (IBCN) research group. As a member of this research group, he is also affiliated with iMinds research institute. He is actively working on the development of an adaptive real-time software defined radio platform based on system on chip components. His main research includes wireless communications, software defined radio and cognitive radio, hardware-software co-design for wireless communications and future networks.

**Christophe Van Praet** received a M.Sc. degree in electro-technical engineering and a Ph.D. degree from Ghent University, in July 2008 and January 2014 respectively. From 2008 to 2014, he was a researcher of INTEC Design, a research group of the Department of Information Technology (INTEC) of Ghent University, affiliated with IMEC. In 2014, he joined iMinds/Ghent University as a postdoctoral researcher. His main research interests include high-speed opto-electronic circuits, with emphasis on CDR techniques for optical access networks, low-power mixed-signal integrated circuit design for telecommunication applications and energy efficient network protocols.

**Merima Kulin** received her M.Sc. degree with honors in Electrical Engineering from the University of Sarajevo, in 2012. In 2013 she joined JSC Elektroprivreda, where she was working as a Data Center Administrator. Since January 2015, she is a PhD student at the department of Information Technology (INTEC) at Ghent University. Within this department, she is working at the Internet Based Communication Networks and Services (IBCN) research group. As a member of this research group, she is also affiliated with the iMinds research institute. Her main research interests include machine learning, network operational intelligence, wireless networks, future internet, self-learning networks and next-generation network architectures.

**Pieter Willemen** received his M.Sc. degree in Industrial Sciences: Intelligent Electronics with Magna cum Laude at the International University College of Leuven in 2013, specializing in FPGA development in the field of wireless communications. Right after, he started working as and currently continues to be a PhD student at the department of Information Technology (INTEC) at Ghent University. Within this department, he is working at the Internet Based Communication Networks and Services (IBCN) research group. As a member of this research group, he is also affiliated with the iMinds research institute. He was a full time contributor to the EU-FP7 SEMAFOUR project, which finished in August 2015 and was applauded by its reviewers for achieving excellent progress and exceeding expectations. His research interests include LTE/WLAN integration, wireless network management, SDR and hardware-software co-design.

**Ingrid Moerman** received her degree in Electrical Engineering (1987) and the Ph.D degree (1992) from the Ghent University, where she became a part-time professor in 2000. She is a staff member of the research group on Internet-Based Communication Networks and Services, IBCN (www.ibcn.intec.ugent.be), where she is leading the research on mobile and wireless communication networks. Since 2006 she joined iMinds, where she is coordinating several interdisciplinary research projects. Her main research interests include: Internet of Things, Cooperative and Cognitive Networks, Wireless Access, and Experimentally-supported research. Ingrid Moerman has a longstanding experience in running national and EU research funded projects. At the European level, Ingrid Moerman is inInternet Research and Experimentation) and Future Networks. Ingrid Moerman is author or co-author of more than 600 publications in international journals or conference proceedings.